%% file: mse_paper.tex
\documentclass[12pt]{article}
\usepackage{amsmath}
\usepackage{graphicx}
\usepackage[dvipsnames]{xcolor}
\usepackage{enumerate}
\usepackage{natbib}
\usepackage{bm}
\usepackage{url} 

\newcommand{\blind}{1}

\begin{document}

\def\spacingset#1{\renewcommand{\baselinestretch}%
{#1}\small\normalsize} \spacingset{1}


\if1\blind
{
  \title{\bf Optimal response surface designs in the presence of model contamination}
  \author{Olga Egorova\hspace{.2cm}\\
    Department of Mathematics, King's College London, UK\\
    and \\
    Steven G. Gilmour \\
    Department of Mathematics, King's College London, UK}
  \maketitle
} \fi

\if0\blind
{
  \bigskip
  \bigskip
  \bigskip
  \begin{center}
    {\LARGE\bf Optimal response surface designs in the presence of model contamination}
\end{center}
  \medskip
} \fi

\bigskip
\begin{abstract}
Complete reliance on the fitted model in response surface experiments is risky and relaxing this assumption, whether out of necessity or intentionally, requires an experimenter to account for multiple conflicting objectives. This work provides a methodological framework of a compound optimality criterion comprising elementary criteria responsible for:  (i) the quality of the confidence region-based inference to be done using the fitted model (DP-/LP-optimality); (ii) improving the ability to test for the lack-of-fit from specified potential model contamination in the form of extra polynomial terms; and (iii) simultaneous minimization of the variance and bias of the fitted model parameters arising from this misspecification. The latter two components have been newly developed in accordance with the model-independent `pure error' approach to the error estimation. The compound criteria and design construction were adapted to blocked experiments.
A point-exchange algorithm was employed for searching for nearly optimal designs. The theoretical work is accompanied by one real and one illustrative example to explore the relationship patterns among the individual components and characteristics of the optimal designs, demonstrating the attainable compromises across the competing objectives and driving some general practical recommendations. 
\end{abstract}

\noindent%
\vfill

\newpage
\spacingset{1.8} 
\section{Introduction}
\label{sec:intro}
\input{introduction}

\section{Criteria Accounting for Model Uncertainty}
\label{sec::criteria}
\input{compound_criterion}

\subsection{Example}
\label{subsec::unblocked_example}
\input{compound_examples}

\section{Blocked Experiments}
\label{sec::compound_blocked}
\input{compound_blocked}

\subsection{Example: case study}
\label{subsec::case_study}
\input{blocked_case_study}

\section{Discussion}
\label{sec::discussion}
\input{discussion}

\if1\blind
{
\section*{Acknowledgements}
OE gratefully acknowledges the funding support from Mathematical Sciences at the University of Southampton, where the majority of this work was carried out as a part of her PhD studies. We acknowledge the use of the IRIDIS High Performance Computing Facility and associated support services at the University of Southampton. The work has been completed as a part of the ``Multi-objective optimal design of experiments'' (MOODE) project, under EPSRC grant EP/T021624/1.
} \fi

\bigskip
\begin{center}
{\large\bf SUPPLEMENTARY MATERIAL}
\end{center}

Contains $\mbox{R}$ code used to obtain optimal designs and the resulting designs presented in the manuscript.

\bibliographystyle{agsm}
\bibliography{mse_paper}

\newpage
\section*{Appendix}
\appendix
\input{appendix_example}
\input{appendix_trace_criteria}
\input{appendix_case_study}
\end{document}

%% file: introduction.tex
Experiments are commonly conducted in order to gain understanding of the effects that different process parameters of interest have on one or more outputs. The quantitative measure of these effects allows interpretable conclusions to be made regarding the shape and strength of the relationships between experimental treatment factors and the measured output or response.

Since the exact true nature of that relationship is generally unknown, some form of approximation is needed, and polynomial functions are able to provide any required accuracy for functions from a certain class of differentiability \citep{Rudin1987real}. Response Surface Methodology \citep{Box1951Roy} aims at identifying the optimum output by fitting second-order polynomials. Greater accuracy would require a polynomial of a higher order and, therefore, more experimental effort. 

Whichever the chosen `primary'  model is, planning a controlled intervention relies on the approximating model assumptions in two main - and quite contradictory - directions. On one hand, there is reliance on the model for inferential purposes, which makes it desirable to ensure good precision of model parameters and/or the prediction accuracy of the untested treatment combinations. The corresponding design aims are usually reflected in the use of well-known optimality criteria ($D$-, $A$-, $G$-, $I$-, etc.). On the other hand, treating the chosen model as absolutely correct, especially at the design stage, is at least too optimistic and could even be compromising the credibility of results. Having a particular model also means that at the stage of planning it is highly desirable to include some control over the model lack of fit, both ensuring its detectability and minimizing its effect on the inferences. 

We deal with such duality of model-dependence and accounting for its misspecification by developing compound optimality criteria, each constructed as a weighted combination of individual criterion functions, with two main features.
\begin{enumerate}
	\item Each criterion function corresponds to a specific desirable property:  either accounting for an aim coming from trusting the model or mitigating the effects from its potential misspecification. The relative importance of the components are reflected by the assigned weights -- and we shall examine the performance of the resulting optimal designs in terms of the individual criteria, and explore the role of the allocation of weights.
	\item Criteria related to inference objectives use model-independent internal variance estimation, `pure error' (\cite{GilmourTrinca2012}). This is the most appropriate and sensible strategy for estimating variance in the case of possible model insufficiency. 
 \end{enumerate}

This work is aligned with the concept of a good design, as summarised by \cite{Box1987empirical}, which should ``make it possible to detect lack of fit'' and ``provide an internal estimate of error from replication'', among other properties. Classical designs, like central composite designs \citep{Myers2009}, have been more popular in practice than optimal designs, partly due to their ability to test lack of fit.

We will focus on response surface experiments with a relatively small number of runs, with the fitted model being a polynomial regression. Section \ref{sec::background} provides the background on the modeling, error estimation and fundamental individual criteria. Controlling the lack-of-fit and the bias arising from the model misspecification are introduced in Section \ref{sec::criteria}, where they are combined with the primary model-driven fundamental criteria in compound optimality criteria. Their adaptation to blocked experiments is described in Section \ref{sec::compound_blocked}.
Examples are  presented which examine the results across various optimal designs, details of the constructed criteria and other properties, followed by a discussion in Section \ref{sec::discussion} with the main conclusions and recommendations. 


\section{Model-dependent optimal design}
\label{sec::background}

Assuming a smooth enough relationship between $k$ experimental treatment factors $X_1, \dots, X_k$ $\in$ $\Theta \subset \mbox{R}^{k}$ and the response of interest $H =\eta(X_1,\ldots, X_k)$ $\in$ $\mbox{R}$, observed as $\bm{Y}$, a suitable polynomial model 
\begin{equation}
\label{eq::primary_model}
\bm{Y}=\bm{X\beta}+\bm{\varepsilon}.
\end{equation} 
is chosen to fit data obtained from $n$ experimental runs. Here $\bm{X}$ is the $n\times p$ model matrix, $\bm{Y}$ is the $n\times 1$ vector of responses; $\bm{\beta}$ is the $p\times 1$ vector of parameters corresponding to the model terms and $\bm{\varepsilon}$ are independent normally distributed random error terms with constant variance: $\bm{\varepsilon}\sim \mathcal{N}(\bm{0},\sigma^{2}\bm{I}_{n})$. In practice, responses are often multivariate, but are typically analyzed separately, so that the design implications are the same as for a single response.

Any inference based on building confidence regions and hypothesis testing following the model fitting relies on the error variance estimate $\hat{\sigma}^2$. The most appropriate estimate, and an advantage of well-designed experiments is `pure' error, which is independent of the parametric model and is derived as the mean square error from fitting the full treatment model
\begin{equation}
\label{eq::treatment_model}
\bm{Y}=\bm{X_{t}\mu_{t}}+\bm{\varepsilon_t},
\end{equation} 
where $\bm{X_{t}}$ is the $n\times t$ matrix, in which the $(i,j)^{th}$ element is equal to $1$ if treatment $j$ is applied to the $i^{th}$ unit and $0$ otherwise. In our context, a ``treatment'' is a combination of levels of factors, and there are $t$ such unique combinations applied in the experiment. The elements of the $t$-dimensional vector $\bm{\mu_{t}}$ are the expected responses for each treatment. The vector of errors $\bm{\varepsilon_t}$ comprises the between-unit variation, such that $\mbox{E}(\bm{\varepsilon_t})=\bm{0}$, $\mbox{Var}(\bm{\varepsilon_t})=\sigma^2_t\bm{I}_{n}$.  \cite{GilmourTrinca2012} give a thorough analysis and discussion in favor of estimating the error from the full treatment model, the correctness of which depends only on the minimal assumption of additive treatment and unit effects and not on which function is used to approximate the relationship of interest.

In a completely randomized experimental setup and assuming model (\ref{eq::primary_model}), this error estimate can be obtained from the further decomposition of the residual sum of squares from fitting the polynomial model into the `pure' error and `lack-of-fit' components, so that $\hat{\sigma}^2_{PE}=\mbox{Pure error SS}/(n-t)$, where $t$ is the number of unique treatments and $d=n-t$ is the pure error degrees of freedom, that is the number of replications.

Model-dependency at the stage of experimental planning is reflected in searching for a design that optimizes a criterion that is a function of the design which captures a specific inference-driven objective. For example, among the most well-known ``alphabetic'' optimality criteria, $D$-, $c$- and $L$-optimality and a series of others \citep{Atkinson2007} target the precision of parameter estimators in model (\ref{eq::primary_model}); while others, like $G$- and $I$-optimality, deal with the prediction variance. These criteria depend on an assumed model both directly through the model matrix used to calculate the criterion and indirectly, since these criteria are formulated assuming $\sigma^2$ is known and so are appropriate only for sufficiently large experiments. \citet{GilmourTrinca2012} derived the alternative pure-error based criteria, which guarantee the presence of replicates in the resulting designs and thus remove the indirect model-dependence. Fundamental criteria were formulated for interval-based inferential properties: minimizing the volume of a $100(1-\alpha_{DP})\%$ confidence region for the model parameters ($DP$-optimality) or the mean squared lengths of the $100(1-\alpha_{LP})\%$ confidence intervals for linear functions of the parameters' estimators' variances ($LP$-optimality). Hence $DP$-optimality is equivalent to minimizing $F_{p,d;1-\alpha_{DP}}^{p}\vert(\bm{X'}\bm{X})^{-1}\vert$ and $LP$-optimality is equivalent to minimizing $F_{1,d;1-\alpha_{LP}}\mbox{tr}\{\bm{W}(\bm{X}'\bm{X})^{-1}\}$,
where $F_{df_1,df_2;1-\alpha}$ is the ``upper $\alpha$-point'' of the F-distribution with $df_1$ and $df_2$ numerator and denominator degrees of freedom respectively. 

Combining multiple desirable objectives in the design can be fulfilled through constructing a compound criterion. This concept is based on the notion of design efficiency, which can be defined for any design matrix $X$ and any criterion $C(X)$ as the ratio with respect to the best (without loss of generality, minimum) value achieved by the optimal design. For example, the $DP$-efficiency of design $\bm{X}$ is
\[
\mbox{Eff}_{DP}(\bm{X})=\frac{\vert \bm{X}'_{*}\bm{X}_{*}\vert^{1/p}/F_{p,d_{*};1-\alpha_{DP}}}{\vert \bm{X}'\bm{X}\vert^{1/p}/F_{p,d;1-\alpha_{DP}}},
\]
where $\bm{X}_{*}$ is the $DP$-optimum design with $d_{*}$ pure error degrees of freedom. In this definition the power $1/p$ brings the efficiency to the scale of the ratio of variances of model coefficients $\bm{\beta}_{i}, i=1\ldots p$ \citep[p.~368]{Atkinson2007}. The efficiency value lies between $0$ and $1$ and is equal to $1$ if and only if the design is optimal according to the criterion of interest.

The compound criterion to be maximized among all the possible designs is obtained then as a weighted product of the individual criterion efficiencies $\mbox{Eff}_{1},\ldots, \mbox{Eff}_{m}$ with corresponding weights $\kappa_{1},\ldots ,\kappa_{m}$ (such that $\kappa_{k}>0$ and $\sum_{k=1}^{m}\kappa_{k}=1$), so that we maximize
\begin{equation}
\label{eq::compound}
\mbox{Eff}^{\kappa_{1}}_{1}(\bm{X})\times\mbox{Eff}^{\kappa_{2}}_{2}(\bm{X})\times\ldots\times\mbox{Eff}^{\kappa_{m}}_{m}(\bm{X}).
\end{equation}

The choice of weights is arbitrary, but is driven by the subjective choices of the experimenter and by the relationship between the objectives of the experiment being planned and the interpretation of the fundamental criteria. 

%% file: compound_criterion.tex
Standard design optimality theory is developed under the assumption that the primary model (\ref{eq::primary_model}) provides the best fit for the data: in many real applications this is quite a strong belief, and in reality we need to take into account at least the possibility that some misspecification is present at the planning stage.

In this work we consider the case when the fitted polynomial model with $p$ parameters is nested within a larger model that is assumed, at the stage of planning the experiment, to provide a better approximation. This model is specified as
\begin{equation}
\label{eq::full_model}
\bm{Y}=\bm{X}_p\bm{\beta}_p+\bm{X}_q\bm{\beta}_q+\bm{\varepsilon},
\end{equation}
where $\bm{X}_q$ is an $n\times q$ extension of the primary model matrix containing the extra $q$ terms that we refer to as `potential terms' and that represent the fitted model disturbance, with vector $\bm{\beta}_q$ denoting the corresponding parameters. They are not of any inferential interest and, moreover, not all of them are necessarily estimable. This will be true when the experiment is relatively small, i.e.\ $n<p+q$, the case we mainly consider here, but might also hold for larger experiments. As usual, we assume independent and normally distributed error terms, $\bm{\varepsilon}\sim \mathcal{N}(\bm{0},\sigma^{2}\bm{I}_{n})$. Note that, even though the extended model is believed to potentially better fit the data, it is not necessarily the one that should be used to obtain the estimates of $\sigma^2$ -- model contamination can still be present in this extended model, which is a strong argument for using the model-independent pure error estimate from the full treatment model (\ref{eq::treatment_model}). On the other hand, the full treatment model cannot be used at the design stage to represent model contamination, since its definition depends on the design itself, through the choice of treatments.

\subsection{Lack-of-fit criterion}
\label{section::LoF_criterion}
To quantify the impact of the potential terms, we adopt a Bayesian approach regarding the full model parameters, as was done by \cite{DuMouchel1994}. A diffuse prior is put on the primary terms - with an arbitrary mean and a variance going to infinity, and a normal prior is put on the potential terms $\bm{\beta}_q\sim\mathcal{N}(0,\bm{\Sigma}_{0})$, where the variance is scaled with respect to the error variance: $\bm{\Sigma}_{0}=\sigma^{2}\tau^{2}\bm{I}_{q}$. Following the normality in model (\ref{eq::full_model}), the posterior distribution of the joint vector of  coefficients $\bm{\beta}  = [\bm{\beta}^T_p, \bm{\beta}^T_q]^T$ is multivariate normal \citep{Koch2007introduction}, conditional on $\sigma,^2$ i.e.\ 
\[
\bm{\beta}|\bm{Y}, \sigma^2 \sim \mathcal{N}(\bm{b},\bm{\Sigma}),
\]
where $\bm{b} = \bm{\Sigma X}^T\bm{Y}$, $\bm{\Sigma} = \sigma^{2}[\bm{X}^T\bm{X} + \bm{K}/\tau^{2}]^{-1}$,  $\bm{X}=[\bm{X}_p, \bm{X}_q]$ and
\[
\bm{K} = \begin{pmatrix}
\bm{0}_{p\times p} & \bm{0}_{p\times q}\\
\bm{0}_{q\times p} & \bm{I}_{q\times q}
\end{pmatrix}.
\]

The marginal posterior distribution of $\bm{\beta}_q$ is also multivariate normal with mean $\bm{b}_q$, the last $q$ elements of $\bm{b}$, and covariance matrix given by the bottom right $q \times q $ submatrix of $\bm{\Sigma}$,
\begin{align*}
\bm{\Sigma}_{qq} = \sigma^{2}[(\bm{X}^T\bm{X} +  \bm{K}/\tau^{2})^{-1}]_{[q,q]} &= 
\sigma^{2}\begin{bmatrix}
\bm{X}^T_p\bm{X}_p& \bm{X}^T_p\bm{X}_q \\
\bm{X}^T_q\bm{X}_p& \bm{X}^T_q\bm{X}_q+\bm{I}_{q}/\tau^{2}
\end{bmatrix}^{-1}_{[q,q]}\\&=
\sigma^{2}[\bm{X}^T_q\bm{X}_q+\bm{I}_{q}/\tau^{2}-\bm{X}^T_q\bm{X}_p(\bm{X}^T_p\bm{X}_p)^{-1}\bm{X}^T_p\bm{X}_q]^{-1}\\&=\sigma^{2}\left[\bm{L}+\bm{I}_{q}/\tau^{2}\right]_,^{-1}
\end{align*} 
where $\bm{L} = \bm{X}^T_q\bm{X}_q-\bm{X}^T_q\bm{X}_p(\bm{X}^T_p\bm{X}_p)^{-1}\bm{X}^T_p\bm{X}_q$ is known in the model-sensitivity design literature as the ``dispersion matrix'' (e.g.\  \cite{Goos2005model}). Its elements provide a measure of the magnitude of the potential terms and how close they are to the orthogonal (residual) subspace defined by the column vectors of the primary model matrix $\bm{X}_p$. 

In the framework of confidence interval- and hypothesis testing-based inference, improving the detectability of the primary model's lack-of-fit in the direction of the potential terms is translated into a criterion function of the design by utilizing the posterior distribution for $\bm{\beta}_q$ derived above and constructing a $100(1-\alpha_{LoF})\%$ posterior credible region which depends on the model matrices and the variance estimate $s^2$ on $\nu$ degrees of freedom \citep{Draper1998}, given by
\begin{equation}
\label{eq::noncentrality}
(\bm{\beta}_{q}-\bm{b}_{q})^{'}(\bm{L}+\bm{I}_{q}/\tau^{2})(\bm{\beta}_{q}-\bm{b}_{q})\leq qs^{2}F_{q,\nu;1-\alpha_{LoF}.}
\end{equation}
Minimizing the volume of this credible region is equivalent to minimizing 
\begin{equation}
\label{eq::LoFDP_criterion}
\left|\bm{L}+\bm{I}_{q}/\tau^{2}\right|^{-1/q}F_{q,d;1-\alpha_{LoF},}
\end{equation}  
and we refer to this as the ``Lack-of-fit DP-criterion''. It is directly related to: (i) the lack-of-fit component in the Generalized $D$-optimality developed by \cite{Goos2005model}, where the residual number of degrees of freedom $\nu$ does not depend on the design; and (ii) $DP$-optimality \citep{GilmourTrinca2012}, with the F-quantile preserved from $\nu = d$ being the number of pure error degrees of freedom in the design. The expression in (\ref{eq::noncentrality}) is related to the non-centrality parameter of the lack-of-fit sum of squares for the primary model, $\delta = \bm{\beta}_{q}^{'}\bm{L}\bm{\beta}_{q}$; maximizing which ($T$-optimality) maximizes the power of the F-test for the primary model lack-of-fit in the direction of the potential terms \citep{Atkinson1975Design}. 

\subsection{MSE-based criterion}
Together with assessing the model contamination, it is also desirable to ``protect" the quality of inference that is to be drawn through fitting the primary model, from the potential presence of extra terms which are not in that model. 

From this point of view, the bias of the parameters' estimators $\hat{\bm{\beta}}_p$ would be of substantial interest; a natural way of evaluating the quality of these estimators is the matrix of  mean squared error \citep{FedorovMontepiedra1997}, which is the $\mathcal{L}_2$-distance between the true and estimated parameter values with respect to the probability distribution measure of $\bm{Y}$ under the assumption of model (\ref{eq::full_model}):
\begin{align}
\label{eq::MSE}
\mbox{MSE}(\bm{\hat{\beta}}_p|\bm{\beta})=&\mathtt{E}_{\bm{Y}|\bm{\beta}}[(\bm{\hat{\beta}}_p-\bm{\beta}_p)(\bm{\hat{\beta}}_p-\bm{\beta}_p)^T]\notag\\=&\sigma^2(\bm{X}_p^T\bm{X}_p)^{-1}+\bm{A}\bm{\beta}_q\bm{\beta}_q^T\bm{A}^T, 
\end{align}
where $\bm{A}=(\bm{X}_p^{T}\bm{X}_p)^{-1}\bm{X}_p^{T}\bm{X}_q$ denotes the $p \times q$ alias matrix, whose elements reflect the measure of the linear relationship between the primary (rows) and potential (columns) terms. 

We start by constructing the determinant-based criterion that would correspond to the overall simultaneous minimization of the mean squared errors by taking the exponential of the average log-determinant of the MSE matrix, across the prior distribution for $\bm{\beta}$, that is we minimize
\begin{equation}
\label{eq::MSE_det}
\exp\{\mathtt{E}_{\bm{\beta}}\log(\det[\mbox{MSE}(\bm{\hat{\beta}}_p|\bm{\beta})])\}.
\end{equation}

Denoting $\bm{M}=\bm{X}_p^{T}\bm{X}_p$ and $\bm{\tilde{\beta}}_q=\bm{\beta}_q/\sigma,$ the determinant and its logarithm in (\ref{eq::MSE_det}) can be decomposed as
\begin{align*}
\det[\mbox{MSE}(\bm{\hat{\beta}}_p|\bm{\beta})]\notag &=\det[\sigma^2\bm{M}^{-1}+\bm{M}^{-1}\bm{X}_p^{T}\bm{X}_q\bm{\beta}_q\bm{\beta}_q^T\bm{X}_q^{T}\bm{X}_p\bm{M}^{-1}]\notag\\&=\sigma^{2p}\det[\bm{M}^{-1}+\bm{M}^{-1}\bm{X}_p^{'}\bm{X}_q\bm{\tilde{\beta}}_q\bm{\tilde{\beta}}_q^T\bm{X}_q^{T}\bm{X}_p\bm{M}^{-1}].\notag
\end{align*}
The matrix determinant lemma \citep[p.~417]{Harville2006matrix} states that, for an invertible matrix $\bm{A}$ and column vectors $\bm{u}$ and $\bm{v}$, given dimension compatibility, 
\[
\det(\bm{A} + \bm{uv}') = \det(\bm{A})(1+\bm{v}'\bm{A}^{-1}\bm{u}).    
\]
Setting $\bm{A} = \bm{M}^{-1}$ and $\bm{u} = \bm{v} = \bm{M}^{-1}\bm{X}_p^{'}\bm{X}_q\bm{\tilde{\beta}}_q$, the determinant above and its logarithm become
\[
\det[\mbox{MSE}(\bm{\hat{\beta}}_p|\bm{\beta})]\notag
=\sigma^{2p}\det[\bm{M}^{-1}](1+\bm{\tilde{\beta}}_q^T\bm{X}_q^{T}\bm{X}_p\bm{M}^{-1}\bm{X}_p^{T}\bm{X}_q\bm{\tilde{\beta}}_q)
\]
and
\[
\log(\det[\mbox{MSE}(\bm{\hat{\beta}}_p|\bm{\beta})])\notag =p\log\sigma^2+\log(\det[\bm{M}^{-1}])+\log(1+\bm{\tilde{\beta}}_q^T\bm{X}_q^{T}\bm{X}_p\bm{M}^{-1}\bm{X}_p^{T}\bm{X}_q\bm{\tilde{\beta}}_q). 
\]
The first summand does not depend on the design, so it will not be included in the criterion; the second summand is the $D$-optimality criterion function, which reflects the variance of estimation; the third summand reflects the bias in estimation. Bringing the criterion in (\ref{eq::MSE_det}) to the scale of a single parameter, the elementary ``MSE(D)-criterion'' is to minimize
\begin{equation}
\label{eq::MSE_D}
\vert\bm{X}_p^{T}\bm{X}_p\vert^{-1/p}\exp\{\mathtt{E}_{\bm{\tilde{\beta}}_q}\log(1+\bm{\tilde{\beta}}_q^T\bm{X}_q^{T}\bm{X}_p(\bm{X}_p^{T}\bm{X}_p)^{-1}\bm{X}_p^{T}\bm{X}_q\bm{\tilde{\beta}}_q)\}^{1/p}.
\end{equation}

Due to the obvious lack of information regarding $\bm{\tilde{\beta}}_q$, the expectation in the second term needs to be evaluated numerically. Expressing the prior variance of $\bm{\beta}_q \sim \mathcal{N}(\bm{0},\tau^{2}\sigma^{2}\bm{I}_{q})$ as a scaled error variance means that $\bm{\tilde{\beta}}_q \sim \mathcal{N}(\bm{0},\tau^{2}\bm{I}_{q})$, and that, quite conveniently, its prior distribution does not depend on the unknown $\sigma^2.$ Then a regular Monte-Carlo sample can be used to evaluate that term: drawing a sample of large size $N$ from the prior, and approximating the expectation above by the average across the sampled values of $\bm{\tilde{\beta}}_{q_i},$ $i=1,\dots, N$, we obtain
\[
\mathtt{E}_{\bm{\tilde{\beta}}_q}\log(1+\bm{\tilde{\beta}}_q^T\bm{X}_q^{T}\bm{X}_p\bm{M}^{-1}\bm{X}_p^{T}\bm{X}_q\bm{\tilde{\beta}}_q) \approx \frac{1}{N}\sum_{i=1}^{N}\log(1+\bm{\tilde{\beta}}_{q_i}^T\bm{X}_q^{T}\bm{X}_p\bm{M}^{-1}\bm{X}_p^{T}\bm{X}_q\bm{\tilde{\beta}}_{q_i}).
\]

One of the alternatives to this computationally demanding approach is to use a point prior for $\bm{\beta}_q$, that is setting  $\bm{\beta}_q = \pm\sigma\tau\bm{1}_q$, where $\bm{1}_q$ is a $q$-dimensional vector of $1$s. Without loss of generality, we shall use $\bm{\beta}_q = \sigma\tau\bm{1}_q$ -- the standard deviation of the initial normal prior, and $\bm{\tilde{\beta}}_q = \tau\bm{1}_q$ with probability $1$, and
\begin{align*}
\mathtt{E}_{\bm{\tilde{\beta}}_q}\log(1+\bm{\tilde{\beta}}_q^T\bm{X}_q^{T}\bm{X}_p\bm{M}^{-1}\bm{X}_p^{T}\bm{X}_q\bm{\tilde{\beta}}_q)  &\approx \log(1+\tau^2\bm{1}^T_q\bm{X}_q^{T}\bm{X}_p\bm{M}^{-1}\bm{X}_p^{T}\bm{X}_q\bm{1}_q) \\&=\log(1+\tau^2\sum_{i,j=1}^{q}[\bm{X}_q^{T}\bm{X}_p\bm{M}^{-1}\bm{X}_p^{T}\bm{X}_q][i,j]),
\end{align*}
with summation of the matrix elements taking considerably less computational time compared with averaging over the prior, even for large $q$.

\subsection*{Compound criteria}

The compound criterion in a general form is constructed to account for the three main inferential objectives: precision of the primary model parameters, identifiability of the lack-of-fit and minimizing the inferential bias from the potential contamination, and is built up as a weighted product of efficiencies (\ref{eq::compound}) with respect to the DP-criterion, lack-of-fit function (\ref{eq::LoFDP_criterion}) and the MSE-based criterion (\ref{eq::MSE_D}).  

The intercept is a nuisance parameter, and so the criteria are adapted in such a way that the full information matrix in the DP criterion and in the first part of the MSE-based component is replaced by the one excluding the intercept $\bm{M}_0 = \bm{X}^T_{p-1}\bm{Q}_0\bm{X}_{p-1}$, where $\bm{X}_{p-1}$ is the primary model matrix without the intercept and $\bm{Q}_0=\bm{I}_n-\frac{1}{n}\bm{1_n1_n}'$. Otherwise obtaining the $MSE(D)$-based individual criterion remains the same, and the compound criterion function that has been amended according to the intercept exclusion is referred to as MSE-DP$_S$ (similar to DP$_S$ from \cite{GilmourTrinca2012}). Then the full determinant-based compound criterion is to minimize
\begin{align}
\label{eq::MSE_DP}
&\left[\left|\bm{X}^T_{p-1}\bm{Q}_0\bm{X}_{p-1}\right|^{-1/(p-1)}F_{p-1,d;1-\alpha_{DP}}\right]^{\kappa_{DP}} \times \notag \\ &\left[\left|\bm{L}+\frac{\bm{I}_{q}}{\tau^{2}}\right|^{-1/q}F_{q,d;1-\alpha_{LoF}}\right]^{\kappa_{LoF}}\times \\ & \left[|\bm{X}'_{p-1}\bm{Q}_0\bm{X}_{p-1}|^{-1}\exp\left(\mathtt{E}_{\bm{\tilde{\beta}}_q}\log(1+\bm{\tilde{\beta}}_q^T\bm{X}_q^{T}\bm{Q}_0\bm{X}_{p-1}\bm{M}_0^{-1}\bm{X}_{p-1}^{T}\bm{Q}_0\bm{X}_q\bm{\tilde{\beta}}_q) \right)\right]_.^{\kappa_{MSE}/(p-1)} \notag
\end{align}

This compound criterion is referred to as the ``compound MSE-DPs-criterion'', where $\alpha_{DP}$ and $\alpha_{LoF}$ denote the confidence levels for the confidence regions for primary $\bm{\beta}_p$ and potential coefficients $\bm{\beta}_q$ in the DP$_S$- and LoF(DP) elementary criteria  respectively. As before, non-negative weights $\kappa_i$ that sum up to $1$ define the compound criterion and are often chosen to reflect the experimenter's priorities. 

Similarly to the determinant-based criteria derived above, we define trace-based Lack-of-fit LP-criterion, MSE(L)-criterion and compound MSE-LP$_S$-criterion in Appendix \ref{appendix_trace_criteria}.

%% file: compound_examples.tex
To explore the practical aspect of applying the the compound criteria, we will study the designs which are optimal in terms of (\ref{eq::MSE_DP}) in the framework of a factorial experiment with $5$ factors, each at three levels. The relatively small number of runs ($n=40$) allows estimation of the full second-order polynomial model ($p=21$), but we assume that the extended model, potentially providing a better fit,  contains also third-order terms: linear-by-linear-by-linear and quadratic-by-linear interactions, so that there are $q=30$ of them. 

We obtain two sets of optimal designs, using two values of the variance scaling parameter $\tau^2=1$ and $\tau^2=1/q$, for each compound criterion. Their properties are summarized in Table \ref{tab::MSE(D)_ex1}; designs optimal with respect to compound MSE-LP$_S$ criteria are presented in Table \ref{tab::MSE(L)_ex1}, Appendix \ref{appendix_trace_example}. 
Every row corresponds to a design that has been obtained as optimal according to the compound criterion defined by the combination of weights $\kappa_i$. We explore the distribution of degrees of freedom between the pure error and lack-of-fit components in the designs and the optimal designs' efficiencies with respect to the individual criteria that are given in the last columns. 


\begin{table}
    \caption{\label{tab::MSE(D)_ex1}Properties of $MSE-DP_{S}$-optimal designs}
    
\fbox{%
\resizebox{\textwidth}{!}{
\begin{tabular}{*{12}{r}}
& \multicolumn{3}{l}{\textbf{Criteria, $\bm{\tau^2=1}$}} & \multicolumn{2}{l}{\textbf{DoF}} & \multicolumn{6}{l}{\textbf{Efficiency,\%}} \\
& \textbf{DP}       & \textbf{LoF(DP)}    & \textbf{MSE(D)}   & \textbf{PE}        & \textbf{LoF}        & \textbf{DP}   & \textbf{LoF(DP)}   & \textbf{MSE(D)}  &  \textbf{LP}       & \textbf{LoF(LP)}   & \textbf{MSE(L)}  \\
1 & 1    & 0    & 0    & \multicolumn{1}{|r}{18} & \multicolumn{1}{r|}{1}  & 100.00 & 47.77  & 91.05  & \multicolumn{1}{|r}{96.38} & 94.92 & 10.70 \\
2 & 0    & 1    & 0    & \multicolumn{1}{|r}{8}  & \multicolumn{1}{r|}{11} & 43.70  & 100.00 & 54.74  & \multicolumn{1}{|r}{0.75}  & 89.99 & 2.08  \\
3 & 0    & 0    & 1    & \multicolumn{1}{|r}{0}  & \multicolumn{1}{r|}{19} & 0.00   & 0.00   & 100.00 & \multicolumn{1}{|r}{0.00}  & 0.00  & 22.32 \\
4 & 0.5  & 0.5  & 0    & \multicolumn{1}{|r}{11} & \multicolumn{1}{r|}{8}  & 78.50  & 87.61  & 88.56  & \multicolumn{1}{|r}{73.88} & 98.85 & 17.66 \\
5 & 0.5  & 0    & 0.5  & \multicolumn{1}{|r}{15} & \multicolumn{1}{r|}{4}  & 97.26  & 56.51  & 93.77  & \multicolumn{1}{|r}{97.55} & 50.04 & 12.74 \\
6 & 0    & 0.5  & 0.5  & \multicolumn{1}{|r}{8}  & \multicolumn{1}{r|}{11} & 64.72  & 96.84  & 87.53  & \multicolumn{1}{|r}{57.04} & 36.17 & 29.33 \\
7 & 1/3  & 1/3  & 1/3  & \multicolumn{1}{|r}{10} & \multicolumn{1}{r|}{9}  & 79.45  & 84.14  & 93.23  & \multicolumn{1}{|r}{81.06} & 43.42 & 16.71 \\
8 & 0.5  & 0.25 & 0.25 & \multicolumn{1}{|r}{13} & \multicolumn{1}{r|}{6}  & 93.38  & 64.35  & 95.55  & \multicolumn{1}{|r}{95.76} & 48.58 & 14.77 \\
9 & 0.25 & 0.5  & 0.25 & \multicolumn{1}{|r}{9} & \multicolumn{1}{r|}{10} & 69.52  & 95.76  & 87.36  & \multicolumn{1}{|r}{63.13} & 40.41 & 25.46 \\ 
10 & 0.25 & 0.25 & 0.5 & \multicolumn{1}{|r}{11} & \multicolumn{1}{r|}{8} & 84.37  & 77.82  & 95.82  & \multicolumn{1}{|r}{87.28} & 98.17 & 16.43 \\
\hline
 & & & & & & & & & & & \\
  & \multicolumn{3}{l}{\textbf{Criteria, $\bm{\tau^2=1/q}$}} & \multicolumn{2}{l}{\textbf{DoF}} & \multicolumn{6}{l}{\textbf{Efficiency,\%}}              \\
  & \textbf{DP}       & \textbf{LoF(DP)}    & \textbf{MSE(D)}   & \textbf{PE}        & \textbf{LoF}        & \textbf{DP}   & \textbf{LoF(DP)}   & \textbf{MSE(D)}  & \textbf{LP}       & \textbf{LoF(LP)}   & \textbf{MSE(L)}  \\
1 & 1    & 0    & 0    & \multicolumn{1}{|r}{18} & \multicolumn{1}{r|}{1}  & 100.00 & 94.41  & 90.60  & \multicolumn{1}{|r}{96.42} & 98.36  & 44.94 \\
2 & 0    & 1    & 0    & \multicolumn{1}{|r}{16} & \multicolumn{1}{r|}{3}  & 39.66  & 100.00 & 37.95  & \multicolumn{1}{|r}{0.13}  & 100.00 & 0.12  \\
3 & 0    & 0    & 1    & \multicolumn{1}{|r}{0}  & \multicolumn{1}{r|}{19} & 0.00   & 0.00   & 100.00 & \multicolumn{1}{|r}{0.00}  & 0.00   & 77.97 \\
4 & 0.5  & 0.5  & 0    & \multicolumn{1}{|r}{18} & \multicolumn{1}{r|}{1}  & 100.00 & 94.41  & 90.60  & \multicolumn{1}{|r}{96.42} & 98.36  & 44.94 \\
5 & 0.5  & 0    & 0.5  & \multicolumn{1}{|r}{17} & \multicolumn{1}{r|}{2}  & 97.31  & 91.28  & 93.98  & \multicolumn{1}{|r}{96.21} & 94.32  & 50.53 \\
6 & 0    & 0.5  & 0.5  & \multicolumn{1}{|r}{15} & \multicolumn{1}{r|}{4}  & 96.10  & 92.31  & 93.29  & \multicolumn{1}{|r}{99.48} & 95.09  & 57.05 \\
7 & 1/3  & 1/3  & 1/3  & \multicolumn{1}{|r}{18} & \multicolumn{1}{r|}{1}  & 100.00 & 94.41  & 90.60  & \multicolumn{1}{|r}{96.42} & 98.36  & 44.94 \\
8 & 0.5  & 0.25 & 0.25 & \multicolumn{1}{|r}{18} & \multicolumn{1}{r|}{1}  & 100.00 & 94.41  & 90.60  & \multicolumn{1}{|r}{96.42} & 98.36  & 44.94 \\
9 & 0.25 & 0.5  & 0.25 & \multicolumn{1}{|r}{18} & \multicolumn{1}{r|}{1}  & 99.96  & 94.33  & 90.65  & \multicolumn{1}{|r}{96.24} & 98.31  & 44.84 \\
10 & 0.25 & 0.25 & 0.5 & \multicolumn{1}{|r}{16} & \multicolumn{1}{r|}{3} & 98.13 & 92.75 & 92.53 & \multicolumn{1}{|r}{96.32} & 96.09 & 47.69\\
\\
\end{tabular}
}}                
\end{table}

Optimal designs were obtained using a point exchange algorithm (\cite{Fedorov1972theory}), with $500$ random starts; the $MSE(D)$-part of the compound criterion was estimated using MC sampling, and this is the most time-consuming part of the computations. When this creates too great a challenge in computing time, we recommend the previously mentioned alternative of using the point prior values of $\bm{\tilde{\beta}_{q}}$. The resulting losses in the efficiencies are quite small, and time savings are substantial - an illustration using this example is given in the author's thesis. 

The resulting designs have more degrees of freedom allocated to the pure error, especially the $MSE-DP_{S}$-optimal designs with $\tau^2=1/q$. The $DP$-optimal design (\#$1$ in both Tables, the design itself can be found in Appendix \ref{appendix::example1}) is also optimal with respect to weight combinations \#$4$, \#$7$ and \#$8$ for smaller $\tau^2$, and performs well in terms of the $MSE(D)$-components. However, its $LoF(DP)$-efficiency drops by roughly half when the scaling parameter goes from $1/q$ to $1$. Also, $LoF(DP)$-optimal designs provide the lowest $DP$-efficiency values (around $40\%$) for any value of $\tau^2$. We see that there is a conflict between the objectives of performing inference on the primary model and testing for lack of fit of that model - no design is highly efficient for both, especially when the lack of fit is expected to be substantial.

Despite the observed conflict between the components, it seems possible to find compromise designs that would be reasonably efficient with respect to all of the components. In practice, we would suggest trying a few combinations of weights and values of $\tau^2$ -- within the experimenter's time restrictions -- which would provide a better insight into the optimal designs in a specific practical problem.

%% file: compound_blocked.tex
In some experiments, where the number of runs is relatively large, and/or the variability between units is high, experimental units are arranged in blocks such that within each block the units are expected to give similar responses. 
Such a restriction to the randomization contributes to 
controlling the variability by separating variation coming from the difference between blocks and the variability between experimental units within the blocks (\cite{Bailey2008design, Grossmann2021partially}).

Under the assumption of additivity of $b$ fixed block effects the polynomial model can be written as
\begin{equation}
\label{eq::blocked_model}
\bm{Y}=\bm{Z\beta}_B+\bm{X_p\beta_p}+\bm{\varepsilon},
\end{equation}
which, in addition to $p$ polynomial terms in $\bm{X}_p$ (excluding the intercept) is comprised of $\bm{Z}$ -- the $n\times b$ matrix, with $(i,j)^{th}$ element equal to $1$ if unit $i$ is in block $j$ and to $0$ otherwise, and $\bm{\beta}_B$, the vector of block effects. 

The full information matrix has the form
\[
\bm{M_B}=
\begin{pmatrix}
\bm{Z}'\bm{Z} & \bm{Z}'\bm{X}_p\\
\bm{X}_p'\bm{Z} & \bm{X}_p'\bm{X}_p
\end{pmatrix}.
\]
Using the rules of inverting blocked matrices \citep[p.~99]{Harville2006matrix}, we can isolate the variance of the polynomial coefficients' estimators as
\begin{align*}
\mbox{Var}(\hat{\bm{\beta}})&=\sigma^2(\bm{M_B}^{-1})_{22}=\sigma^2(\bm{X}_p'\bm{QX}_p)^{-1},
\end{align*}
where $\bm{Q}=\bm{I}-\bm{Z}(\bm{Z}'\bm{Z})^{-1}\bm{Z}'$.
The $DP$- and $LP$-criteria are best replaced by $DP_S$ and $LP_S$ in the context of a blocked experiment, and these can be straightforwardly defined as minimizing
\begin{equation}
\label{eq::DPs_blocked}
(F_{p,d_B;1-\alpha_{DP}})^{p}\vert (\bm{X}_p'\bm{Q}\bm{X}_p)^{-1}\vert
\end{equation}
and
\begin{equation}
\label{eq::LPs_blocked}
F_{1,d_B;1-\alpha_{LP}}\mbox{trace}\{\bm{W}(\bm{X}_p'\bm{Q}\bm{X}_p)^{-1}\}
\end{equation}
respectively. The number of pure error degrees of freedom is now calculated as $d_B=n-\mbox{rank}[\bm{Z}:\bm{T}]$, where $\bm{T}$ is the $n\times t$ matrix whose elements indicate the treatments (\citealp{GilmourTrinca2012}), providing the number of replications minus those taken for the estimation of block contrasts.  

To adapt the derivation of the lack-of-fit and MSE-based criteria to the blocked experiments, we start by formulating the model comprised of both primary terms and possible contamination in the form of potential terms, now for blocked experiments, giving
\[
\bm{Y}=\bm{Z\beta}_{B}+\bm{X}_{p}\bm{\beta}_{p}+\bm{X}_{q}\bm{\beta}_{q}+\bm{\varepsilon}.
\]
Denote the $n\times(b+p)$ model matrix of the block and primary terms by $\tilde{\bm{X}}_{p}=[\bm{Z},\bm{X}_{p}]$ and let $\bm{\tilde{\beta}}_p=[\bm{\beta}_{B},\bm{\beta}_{p}]'$ be the joint vector of fixed block effects and primary model terms, and by $\bm{\hat{\tilde{\beta}}}_p$ we denote the vector of the corresponding estimates. It is worth noting that the number of primary terms $p$ does not include the intercept, as it is aliased with the block effects. 

\subsection{Lack-of-fit criteria}
The information matrix for model (\ref{eq::blocked_model}), up to a multiple of $1/\sigma^2$, is: 
\[
\bm{M_B}=
\begin{pmatrix}
\bm{Z}^T\bm{Z} & \bm{Z}^T\bm{X}_{p} & \bm{Z}^T\bm{X}_{q}\\
\bm{X}^T_{p}\bm{Z} & \bm{X}^T_{p}\bm{X}_{p} & \bm{X}^T_{p}\bm{X}_{q}\\
\bm{X}^T_{q}\bm{Z} & \bm{X}^T_{q}\bm{X}_{p} & \bm{X}^T_{q}\bm{X}_{q}+\bm{I}_{q}/\tau^2
\end{pmatrix}
=
\begin{pmatrix}
\bm{\tilde{X}}^T_{p}\bm{\tilde{X}}_{p} & \bm{\tilde{X}}^T_{p}\bm{X}_{q}\\
\bm{X}^T_{q}\bm{\tilde{X}}_{p} & \bm{X}^T_{q}\bm{X}_{q}+\bm{I}_{q}/\tau^2
\end{pmatrix}_{.}
\]

Assuming the same normal prior on $\bm{\beta}_q$ $\sim \mathcal{N}(\bm{0}, \tau^2\sigma^2\bm{I}_q)$ as for completely randomized designs, we can construct the variance-covariance matrix corresponding to the potential terms, which would be the lower right submatrix of the inverse of $\bm{M_B}$: $\bm{\tilde{\Sigma}}_{qq}=\sigma^2[\bm{M}^{-1}_B]_{22}$, i.e.\ 
\begin{align*}
\bm{\tilde{\Sigma}}_{qq}&=\sigma^2([\bm{M_B}]_{22}-[\bm{M_B}]_{21}([\bm{M_B}]_{11})^{-1}[\bm{M_B}]_{12})^{-1}\\
&=\sigma^2(\bm{X}^T_{q}\bm{X}_{q}+\bm{I}_{q}/\tau^2-\bm{X}^T_{q}\bm{\tilde{X}}_{p}(\bm{\tilde{X}}^T_{p}\bm{\tilde{X}}_{p})^{-1}\bm{\tilde{X}}^T_{p}\bm{X}_{q})^{-1}\\&=\sigma^2\left(\bm{\tilde{L}}+\frac{\bm{I}_{q}}{\tau^2}\right), \mbox{ where }\bm{\tilde{L}}=\bm{X}^T_{q}\bm{X}_{q}-\bm{X}^T_{q}\bm{\tilde{X}}_{p}(\bm{\tilde{X}}^T_{p}\bm{\tilde{X}}_{p})^{-1}\bm{\tilde{X}}^T_{p}\bm{X}_{q}.
\end{align*}

Therefore, the lack-of-fit criteria in (\ref{eq::LoFDP_criterion}) and (\ref{eq::LoFLP_criterion}) are adjusted for blocked experiments by replacing the primary terms matrix $\bm{X}_{p}$ by the extended matrix $\bm{\tilde{X}}_{p}$ and the dispersion matrix $\bm{L}$ by $\bm{\tilde{L}}$ as obtained above.

\subsection{MSE-based criteria}
\label{sec::mse_blocked}
As for the MSE-based measure of the shift in the primary terms estimates, we first consider the overall mean square matrix  
\begin{align}
\label{eq::MSE_b}
\mbox{MSE}(\bm{\hat{\tilde{\beta}}}_p|\bm{\tilde{\beta}})=&\mathtt{E}_{\bm{Y}|\bm{\beta}}[(\bm{\hat{\tilde{\beta}}}_p-\bm{\tilde{\beta}}_p)(\bm{\hat{\tilde{\beta}}}_p-\bm{\tilde{\beta}}_p)'] \notag\\=&\sigma^2(\bm{\tilde{X}}_p^{'}\bm{\tilde{X}}_p)^{-1}+\bm{\tilde{A}}\bm{\beta}_q\bm{\beta}_q^T\bm{\tilde{A}}^T, 
\end{align}
with $\bm{\tilde{A}}=(\bm{\tilde{X}}_p^T\bm{\tilde{X}}_p)^{-1}\bm{\tilde{X}}_p^{'T}\bm{X}_q$ being the alias matrix, 
and its partition with respect to block and primary effects, to get
\begin{align*}
&\mathtt{E}_{\bm{Y}|\bm{\beta}}[(\bm{\hat{\tilde{\beta}}}_p-\bm{\tilde{\beta}}_p)(\bm{\hat{\tilde{\beta}}}_p-\bm{\tilde{\beta}}_p)^T] \notag\\ 
&=\mathtt{E}_{\bm{Y}|\bm{\beta}}\{[\hat{\tilde{\beta}}_{p1}-\tilde{\beta}_{p1},\ldots,
\hat{\tilde{\beta}}_{pb}-\tilde{\beta}_{pb}, \hat{\tilde{\beta}}_{p b+1}-\tilde{\beta}_{p b+1},\ldots, \hat{\tilde{\beta}}_{p b+p}-\tilde{\beta}_{p b+p}]\times \notag\\ &[\hat{\tilde{\beta}}_{p 1}-\tilde{\beta}_{p1},\ldots,
\hat{\tilde{\beta}}_{pb}-\tilde{\beta}_{pb}, \hat{\tilde{\beta}}_{p b+1}-\tilde{\beta}_{p b+1},\ldots, \hat{\tilde{\beta}}_{p b+p}-\tilde{\beta}_{p b+p}]^T\}\notag\\
&=\mathtt{E}_{\bm{Y}|\bm{\beta}}\{[\bm{\hat{\beta}}_B-\bm{\beta}_B, \bm{\hat{\beta}}_p-\bm{\beta}_p][\bm{\hat{\beta}}_B-\bm{\beta}_B, \bm{\hat{\beta}}_p-\bm{\beta}_p]^T\}\notag\\
&=\begin{bmatrix}
\mathtt{E}_{\bm{Y}|\bm{\beta}}(\bm{\hat{\beta}}_B-\bm{\beta}_B)(\bm{\hat{\beta}}_B-\bm{\beta}_B)^T & \mathtt{E}_{\bm{Y}|\bm{\beta}}(\bm{\hat{\beta}}_B-\bm{\beta}_B)(\bm{\hat{\beta}}_p-\bm{\beta}_p)^T\\
\mathtt{E}_{\bm{Y}|\bm{\beta}}(\bm{\hat{\beta}}_p-\bm{\beta}_p)(\bm{\hat{\beta}}_B-\bm{\beta}_B)^T & \mathtt{E}_{\bm{Y}|\bm{\beta}}(\bm{\hat{\beta}}_p-\bm{\beta}_p)(\bm{\hat{\beta}}_p-\bm{\beta}_p)^T
\end{bmatrix}.
\end{align*}

The part  corresponding to the bias of the primary terms $\bm{\beta}_p$ is the lower right $p \times p$ submatrix, and we can extract it from the MSE expression in (\ref{eq::MSE_b}). The respective submatrix of the first summand  is
\[
[\sigma^2(\bm{\tilde{X}}_p^{T}\bm{\tilde{X}}_p)^{-1}]_{22}=\sigma^2(\bm{X}^T_{p}
\bm{QX}_{p})^{-1},
\]
where $\bm{Q}=\bm{I}-\bm{Z}(\bm{Z}^T\bm{Z})^{-1}\bm{Z}^T$. \\
Using the matrix inversion rule for block matrices \citep{Harville2006matrix}, we now consider
\begin{align*}
\bm{\tilde{A}}=&\left(
\begin{bmatrix}
\bm{Z}^T\\
\bm{X}^T_{p}
\end{bmatrix}
\begin{bmatrix}
\bm{Z} & \bm{X}_{p}
\end{bmatrix}\right)^{-1}
\begin{bmatrix}
\bm{Z}^T\\
\bm{X}^T_{p}
\end{bmatrix}\bm{X}_{q}=
\begin{pmatrix}
\bm{Z}^T\bm{Z} & \bm{Z}^T\bm{X}_{p}\\
\bm{X}^T_{p}\bm{Z} & \bm{X}^T_{p}\bm{X}_{p}
\end{pmatrix}^{-1}
\begin{bmatrix}
\bm{Z}^T\\
\bm{X}^T_{p}
\end{bmatrix}\bm{X}_{q}
\\
=&
\begin{bmatrix}
(\bm{Z}^T\bm{PZ})^{-1} & -(\bm{Z}^T\bm{PZ})^{-1}\bm{Z}^T\bm{X}_{p}(\bm{X}^T_{p}\bm{X}_{p})^{-1} \\
-(\bm{X}^T_{p}\bm{QX}_{p})^{-1}\bm{X}^T_{p}\bm{Z}(\bm{Z}^T\bm{Z})^{-1} & (\bm{X}^T_{p}\bm{QX}_{p})^{-1}
\end{bmatrix}
\begin{bmatrix}
\bm{Z}^T\\
\bm{X}^T_{p}
\end{bmatrix}\bm{X}_{q}\\ =&
\begin{bmatrix}
(\bm{Z}^T\bm{PZ})^{-1}\bm{Z}^T\bm{PX}_{q}\\
(\bm{X}^T_{p}\bm{QX}_{p})^{-1}\bm{X}^T_{p}\bm{QX}_{q}
\end{bmatrix},
\end{align*}
where $\bm{P}=\bm{I}-\bm{X}_{p}(\bm{X}^T_{p}\bm{X}_{p})^{-1}\bm{X}^T_{p}$, $\bm{ZZ}'$, $\bm{X}'_{1}\bm{X}_{1}$ and $\bm{Z}'\bm{PZ}$ are all invertible and, therefore, the operations are legitimate. Now denote $\bm{R_P} = (\bm{Z}^T\bm{PZ})^{-1}\bm{Z}^T\bm{PX}_{q}$ and 
$\bm{R_Q} = (\bm{X}^T_{p}\bm{QX}_{p})^{-1}\bm{X}^T_{p}\bm{QX}_{q}$, and
consider the second summand in (\ref{eq::MSE_b}),
\begin{align*}
&\bm{\tilde{A}}\bm{\beta}_q\bm{\beta}_q^T\bm{\tilde{A}}^T=
\begin{bmatrix}
\bm{R_P}\bm{\beta}_q \\
\bm{R_Q}\bm{\beta}_q
\end{bmatrix}
\begin{bmatrix}
\bm{\beta}^T_q\bm{R_P}^T & \bm{\beta}^T_q\bm{R_Q}^T
\end{bmatrix}=
\begin{bmatrix}
\bm{R_P}\bm{\beta}_q\bm{\beta}^T_q\bm{R_P}^T & \bm{R_P}\bm{\beta}_q\bm{\beta}^T_q\bm{R_Q}^T \\
\bm{R_Q}\bm{\beta}_q\bm{\beta}^T_q\bm{R_P}^T & \bm{R_Q}\bm{\beta}_q\bm{\beta}^T_q\bm{R_Q}^T
\end{bmatrix}.
\end{align*}

Then the submatrix of (\ref{eq::MSE_b}) corresponding to the primary terms is
\begin{align}
\label{eq::mesb_submatrix}
\mbox{MSE}(\bm{\hat{\tilde{\beta}}}_p|\bm{\tilde{\beta}})_{pp}=& \sigma^2(\bm{X}^T_{p}\bm{QX}_{p})^{-1}+\bm{R_Q}\bm{\beta}_q\bm{\beta}^T_q\bm{R_Q}^T\notag\\=& \sigma^2\bm{\tilde{M}}^{-1}+\bm{\tilde{M}}^{-1}\bm{X}^T_{p}\bm{QX}_{q}\bm{\beta}_q\bm{\beta}^T_q\bm{X}^T_{q}\bm{QX}_{p}\bm{\tilde{M}},^{-1}
\end{align}
where $\bm{\tilde{M}}=\bm{X}^T_{p}\bm{QX}_{p}.$

As in the unblocked case, we first look at the determinant of the corresponding submatrix (\ref{eq::mesb_submatrix}),
\begin{align}
\label{eq::MSE_B_det}
\det[\mbox{MSE}(\bm{\hat{\tilde{\beta}}}_p|\bm{\tilde{\beta}})_{pp}]=&\det[\sigma^2\bm{\tilde{M}}^{-1}+\bm{\tilde{M}}^{-1}\bm{X}^T_{p}\bm{QX}_{q}\bm{\beta}_q\bm{\beta}^T_q\bm{X}^T_{q}\bm{QX}_{p}\bm{\tilde{M}}^{-1}]\notag\\
=& \sigma^{2p}\det[\bm{\tilde{M}}^{-1}+\bm{\tilde{M}}^{-1}\bm{X}^T_{p}\bm{QX}_{q}\bm{\tilde{\beta}}_q\bm{\tilde{\beta}}^T_q\bm{X}^T_{q}\bm{QX}_{p}\bm{\tilde{M}}^{-1}]\notag\\
=& \sigma^{2p}\det[\bm{\tilde{M}}^{-1}](1+\bm{\tilde{\beta}}^T_q\bm{X}^T_{q}\bm{QX}^T_{p}\bm{\tilde{M}}^{-1}\bm{X}^T_{p}\bm{QX}_{q}\bm{\tilde{\beta}}_q).
\end{align}

The $q$-dimensional random vector $\bm{\tilde{\beta}}_q$, as before, follows $\mathcal{N}(\bm{0},\tau^{2}\bm{I}_{q})$, so that this prior does not depend on the error variance $\sigma^2$. Next, taking the expectation of the logarithm of (\ref{eq::MSE_B_det}) over the prior distribution is identical to the derivations leading to (\ref{eq::MSE_D}). The MSE(D)-component then becomes
\begin{equation}
\label{eq::mse_b_component}
\log(\det[\bm{\tilde{M}}^{-1}])+\mathtt{E}_{\bm{\tilde{\beta}}_2}\log(1+\bm{\tilde{\beta}}_2'\bm{X}_2^{'}\bm{QX}_1\bm{\tilde{M}}^{-1}\bm{X}_1^{'}\bm{QX}_2\bm{\tilde{\beta}}_2),
\end{equation}
and the resulting determinant-based compound criterion for a blocked experiment is to minimize
\begin{align}
\label{eq::MSE_D_B}
&\left[\left|(\bm{X}^T_{p}\bm{Q}\bm{X}_{p})^{-1}\right|^{1/p}F_{p,d_B;1-\alpha_{DP}}\right]^{\kappa_{DP}} \times \notag \\ &\left[\left|\bm{\tilde{L}}+\frac{\bm{I}_{q}}{\tau^{2}}\right|^{-1/q}F_{q,d_B;1-\alpha_{LoF}}\right]^{\kappa_{LoF}}\times \notag \\ & \left[|\bm{X}^T_{p}\bm{QX}_{p}|^{-1}\exp\left(\frac{1}{N}\sum_{i=1}^{N}\log(1+\bm{\tilde{\beta}}_{2i}'\bm{X}_q^{T}\bm{QX}_p\bm{\tilde{M}}^{-1}\bm{X}_p^{T}\bm{QX}_q\bm{\tilde{\beta}}_{2i})\right)\right]^{\kappa_{MSE}/p}.
\end{align}

The probability levels $\alpha_i$ and weights $\kappa_j$ have the same meanings as in the unblocked case and, as was noted earlier, the number of pure error degrees of freedom $d_B$ accounts for the comparisons between blocks.

%% file: blocked_case_study.tex
We shall consider an example of a real-life design problem, the details of which are confidential, and explore a range of solutions provided by the compound optimality criteria.
A company specializing in the production of food supplements for animals were to conduct an experiment to figure out whether a slight decrease in the recommended dosages of three particular products taken together would have a meaningful impact on the resulting ``performance'', which is expressed in terms of some continuous response. The dosage range of interest for each supplement (experimental factor) is from $90\%$ to $100\%$ of the standard recommendation; it is desired that there would be three levels (i.e.\ taking the values of $90\%$, $95\%$ and $100\%$). Carrying out the experiment with more than three levels was more complicated: measuring, for example, $92.5\%$ of the recommended dosage was inconvenient. 

The treatments were to be applied to $n = 36$ cages of animals (experimental units), which would be allocated in $b=2$ equal sized blocks. The primary response surface model would contain all linear, quadratic and bilinear interaction terms ($p=9$). As it was suspected that increasing dosages beyond certain values might not have an impact, it was reasonable to suggest that there might be non-quadratic curvature of the fitted function, meaning that addition of higher order terms would provide a better fit for the data, and it was desirable to accommodate that possibility at the design stage. In the extended model we accounted for $q=10$ potential terms (linear-by-linear-by-linear, quadratic-by-linear and cubic), with the notation the same as before, so that
\[
\bm{Y}=\bm{Z\beta}_{B}+\bm{X}_{p}\bm{\beta}_{p}+\bm{X}_{q}\bm{\beta}_{q}+\bm{\varepsilon}.
\]

The design search was performed among a larger $5$-level candidate set of points, but due to the form of the $2$nd order polynomial primary model and the criteria used, the resulting optimal designs had only $3$ levels. The experimenters also wished to have at least two center points in each block to ensure representation of the conditions thought \emph{a priori} most likely to be best (with dosages of $95\%$ for each supplement), i.e.\ $4$ runs in total were fixed beforehand. This constraint was built directly into the search procedure; we also obtained designs without this restriction and evaluated the efficiency losses.

The experimenters preferred using the determinant-based criterion, since the primary inferential interest was on the overall impact of the model terms, so the search was conducted with respect to the compound MSE-DP-criteria for blocked experiments (\ref{eq::MSE_B_det}). We considered three sets of weights: (1) with the weight being equally distributed among the components; (2) a bit more weight ($0.4$) put on the DP-component, with the rest allocated equally between the lack-of-fit and MSE(D)-components; and (3) with half of the weight on the MSE(D)-component with the rest of it distributed evenly among the others. For each combination of weights we will consider two cases, $\tau^2=1$ and $\tau^2=1/q$. As for the number of Monte Carlo samples used to estimate the third criterion component, for $\tau^2=1$ we set $N=500$, and for $\tau^2=1/q$ we set $N=1000$ in order to have a sufficiently small relative estimation error. The search for each design was performed with $50$ random starts.

Table \ref{tab::MSE(D)_caseCP} contains the summaries of the optimal designs. The two types of efficiencies are presented: with respect to the individual criteria with (``CP Efficiency'') and without (``No CP Efficiency'') the pre-specified two center points per block. The former will be, obviously, larger and the differences represent the magnitude of the losses by restricting the set of designs to be considered. The``Relative Efficiency'' column reflects how well the given design performs with respect to the optimal design in terms of the same compound criterion, obtained without fixing the center points. 

\begin{table}[!h]
    \caption{\label{tab::MSE(D)_caseCP}Case-study: properties of $MSE-DP$-optimal blocked designs, with two center points per block\\}
    
\fbox{%
\scalebox{0.7}{ 
\begin{tabular}{*{12}{r}*{1}{c}}
& \multicolumn{3}{l}{\textbf{Criteria, $\bm{\tau^2=1}$}} & \multicolumn{2}{l}{\textbf{DoF}} & \multicolumn{3}{l}{\textbf{No CP Efficiency,\%}}  & \multicolumn{3}{l}{\textbf{CP Efficiency,\%}}& \multicolumn{1}{l}{\textbf{Relative}}                          \\
   & \textbf{DP}       & \textbf{LoF(DP)}    & \textbf{MSE(D)}   & \textbf{PE}        & \textbf{LoF}        & \textbf{DP}   & \textbf{LoF(DP)}   & \textbf{MSE(D)}  &  \textbf{DP}       & \textbf{LoF(DP)}   & \textbf{MSE(D)} & \textbf{Efficiency,\%} \\
1 & 1/3 & 1/3 & 1/3 & \multicolumn{1}{|r}{14} & \multicolumn{1}{r|}{11} & 88.63 & 90.03 & 99.97 & \multicolumn{1}{|r}{92.89} & 97.59 & \multicolumn{1}{r|}{100.75} & 98.18 \\
2 & 0.4 & 0.2 & 0.4 & \multicolumn{1}{|r}{14} & \multicolumn{1}{r|}{11} & 88.35 & 90.14 & 99.15 & \multicolumn{1}{|r}{92.60} & 97.70 & \multicolumn{1}{r|}{99.92} & 98.31 \\
3 & 0.25 & 0.25 & 0.5 & \multicolumn{1}{|r}{14} & \multicolumn{1}{r|}{11} & 88.63 & 90.03 & 99.75 & \multicolumn{1}{|r}{92.89} & 97.59 & \multicolumn{1}{r|}{100.53} & 98.90 \\
4 & 1 & 0 & 0 & \multicolumn{1}{|r}{20} & \multicolumn{1}{r|}{5} & 95.41 & 62.13 & 95.24 & \multicolumn{1}{|r}{100.00} & 67.34 & \multicolumn{1}{r|}{95.98} & 95.58 \\
5 & 0 & 1 & 0 & \multicolumn{1}{|r}{14} & \multicolumn{1}{r|}{11} & 66.49 & 92.26 & 79.91 & \multicolumn{1}{|r}{69.69} & 100.00 & \multicolumn{1}{r|}{80.53} & 92.26 \\
6 & 0 & 0 & 1 & \multicolumn{1}{|r}{14} & \multicolumn{1}{r|}{11} & 88.31 & 88.35 & 99.23 & \multicolumn{1}{|r}{92.55} & 95.77 & \multicolumn{1}{r|}{100.00} & 99.23 \\ 
\\ \hline
 & & & & & & & & & & & & \\
   & \multicolumn{3}{l}{\textbf{Criteria, $\bm{\tau^2=1/q}$}} & \multicolumn{2}{l}{\textbf{DoF}} & \multicolumn{3}{l}{\textbf{No CP Efficiency,\%}}  & \multicolumn{3}{l}{\textbf{CP Efficiency,\%}}& \multicolumn{1}{l}{\textbf{Relative}}                          \\
   & \textbf{DP}       & \textbf{LoF(DP)}    & \textbf{MSE(D)}   & \textbf{PE}        & \textbf{LoF}        & \textbf{DP}   & \textbf{LoF(DP)}   & \textbf{MSE(D)}  &  \textbf{DP}       & \textbf{LoF(DP)}   & \textbf{MSE(D)} & \textbf{Efficiency,\%} \\
1 & 1/3 & 1/3 & 1/3 & \multicolumn{1}{|r}{18} & \multicolumn{1}{r|}{7} & 92.52 & 95.13 & 95.86 & \multicolumn{1}{|r}{96.97} & 98.00 & \multicolumn{1}{r|}{96.55} & 94.95 \\
2 & 0.4 & 0.2 & 0.4 & \multicolumn{1}{|r}{18} & \multicolumn{1}{r|}{7} & 94.19 & 92.19 & 96.54 & \multicolumn{1}{|r}{98.72} & 94.98 & \multicolumn{1}{r|}{97.24} & 95.34 \\
3 & 0.25 & 0.25 & 0.5 & \multicolumn{1}{|r}{17} & \multicolumn{1}{r|}{8} & 92.44 & 93.70 & 97.05 & \multicolumn{1}{|r}{96.88} & 96.53 & \multicolumn{1}{r|}{97.75} & 95.72 \\
4 & 1 & 0 & 0 & \multicolumn{1}{|r}{20} & \multicolumn{1}{r|}{5} & 95.41 & 90.63 & 95.66 & \multicolumn{1}{|r}{100.00} & 93.37 & \multicolumn{1}{r|}{96.35} & 95.41 \\
5 & 0 & 1 & 0 & \multicolumn{1}{|r}{22} & \multicolumn{1}{r|}{3} & 76.11 & 97.07 & 77.10 & \multicolumn{1}{|r}{79.77} & 100.00 & \multicolumn{1}{r|}{77.65} & 97.07 \\
6 & 0 & 0 & 1 & \multicolumn{1}{|r}{14} & \multicolumn{1}{r|}{11} & 88.37 & 90.37 & 99.29 & \multicolumn{1}{|r}{92.62} & 93.10 & \multicolumn{1}{r|}{100.00} & 99.29\\
\end{tabular}
}}
\end{table}

The main feature observed is that in general individual efficiency values are quite large. This might be attributed to the large number of available residual degrees of freedom ($25$), and this contributes to better compromises achievable among the three criterion components. The imbalance in the distribution of the residual degrees of freedom is not strong, though the prevalence of pure error degrees of freedom is quite consistent. 

When the model disturbance effect is assumed to be quite small ($\tau^2 = 1/q$), the individual efficiencies are larger in general, and the compromise might be more feasible. Relative efficiencies are quite good, losses due to forcing the inclusion of center points among the first three designs (optimal with respect to the compound criteria) do not exceed $1.82\%$ for $\tau^2 = 1$ and $5.05\%$ for $\tau^2=1/q$.

It is notable that designs \#$1$ and \#$3$ (for $\tau^2 = 1$) are the same, and its $MSE(D)$-value is better than of the design \#$6$, which was constructed as being optimal with respect to this component -- illustrating that the algorithm finds nearly optimal designs, but might miss the optimal design. This design was chosen to carry out the experiment; it has been run successfully and useful conclusions were drawn from the data collected. It can be found in Appendix \ref{appendix_case_study}, Figure \ref{fig::CS_design} and Table \ref{tab::CS_Design}. There are only two center points in each block, and replicates of other points are split evenly between blocks (except for the $(-1,1,1)$ point which is duplicated in the first block only).

As for the time costs, on average an optimal design was found in $13-15$ hours, which was acceptable in this particular case. Sometimes, however, it took up to $20-24$ hours, so some extra time allowance should be accounted for when using these criteria and this search algorithm and/or the extensive sampling might be replaced by a less demanding alternative.

%% file: discussion.tex
The possibility of a potentially ``better'' model should not be ignored at the planning stage, and combining the primary inferential individual criteria (DP-, LP-) with the developed  lack-of-fit and MSE-based components result in compound criteria allowing for compromises across competing objectives and tools for decision making reflecting the priorities and aims of the experimentation. 

We combined the component criteria into compound criteria and explored the dynamics in weight allocations and the performances of the optimal designs. Alternatively, one could carry out the multi-objective optimization by constructing a Pareto front of the designs, first introduced by \cite{Lu2011optimization} and developed further and adapted for various criteria and experimental frameworks, e.g. \cite{Cao2017hybrid}. \cite{Sambo2014Coordinate} presented an algorithm for optimizing with respect to $D$- and $I$-optimality in split-plot designs; \cite{Borrotti2016Multi} extended it to the multi-stratum framework and larger number of individual criteria, and later to guarantee the pure error estimation of the variance components \citep{Borrotti2022}. It was also used by \cite{Leonard2017} as Bayesian DP-optimality criterion was introduced in the context of potential terms in screening experiments, and designs assessed in terms of various criteria.

Various forms of model contamination have been studied in the literature: unknown forms considered by e.g. \cite{Notz1989Optimal}, \cite{Wiens1993Designs}, \cite{Woods2005designing}, and T-optimality allowing the choice between two polynomials (\cite{Atkinson1975Design}, \cite{Dette2012T}). \cite{Wiens1992Minimax, Wiens2000Bias, Wiens2009Robust} focused on constructing designs robust against certain classes of  model faults, the Q$_B$ criterion was developed \citep{Tsai2007Three} and generalized \citep{Tsai2010General} to make use of prior model uncertainty knowledge and make inference from a family of nested models. 
\cite{Goos2005model} considered the same framework of model misspecification as we do, and developed generalized criteria combining model-robust and model-sensitive approaches; the authors orthonormalized the primary and potential subspaces \citep{Kobilinsky1998} which was necessary for their bias component derivation and it also ensured the same interpretation of  $\bm{\beta}_p$ in both primary and extended models; we did not perform orthonormalization, since the scaling of the factors' levels and the assumption of the contamination not being too large mean it has little impact, but it is a possible option to consider.

The inferential focus of the work has been on the quality of the fitted model parameters; other objectives reflected in different criteria might also be reviewed following the robust pure-error approach and included in the compound criteria. \cite{de2019prediction} introduced a variety of I-type criteria -- focusing on prediction properties --  and combining them with the D-type criteria. 

Our methodology is fairly flexible in terms of specifying the form and scale of model contamination (which in itself can be seen either as a convenient and situation-appropriate advantage or an additional source of uncertainty), and is straightforward in application to experiments. 
It would be of certain interest to expand the MSE-criteria in particular to more complex structures of treatments and experimental units, for example, crossed structures, networks \citep{Koutra2021optimal} and complex interventions, such as sequential designs (e.g. \cite{Gilmour1995stopping}).

%% file: appendix_example.tex

\section{MSE-DP-optimal completely randomized design}
\label{appendix::example1}
The $DP$-optimal design \#$1$ for the experiment described in Section \ref{subsec::unblocked_example}, which is also optimal: (i) for the uniform weight allocation across the three components; (ii) for the weight equally distributed between the DP- and LoF(DP)-components; and (iii) in terms of the criterion with half weight on the DP-component and a quarter on the LoF(DP) and MSE(D) components (all for $\tau^2 = 1/q$). The design has $18$ pure error degrees of freedom which arise from pairs of replicated points -- the only unreplicated points are $\#3$, $\#24$, $\#33$ and $\#34$. Fourteen of these pairs come from replicates of the $2^{5-1}$ fractional factorial, and the $4$ remaining replicated points have $2$ or $3$ factors set to $0$.

\begin{table}[!h]
    \caption{\label{tab::DP_design}MSE-DP-optimal design}
\centering
\fbox{%
\scalebox{0.8}{
\begin{tabular}{rrrrrr||rrrrrr||rrrrrr||rrrrrr}
1 & -1 & -1 & -1 & -1 & -1 & 11 & -1 & 1 & -1 & -1 & 1 & 21 & 0 & -1 & -1 & 1 & 0 & 31 & 1 & -1 & 1 & 1 & 1 \\
2 & -1 & -1 & -1 & -1 & -1 & 12 & -1 & 1 & -1 & 1 & -1 & 22 & 0 & 0 & 0 & -1 & 1 & 32 & 1 & -1 & 1 & 1 & 1 \\
3 & -1 & -1 & 0 & 1 & 1 & 13 & -1 & 1 & -1 & 1 & -1 & 23 & 0 & 0 & 0 & -1 & 1 & 33 & 1 & 0 & -1 & -1 & 0 \\
4 & -1 & -1 & 1 & -1 & 1 & 14 & -1 & 1 & 0 & 0 & 0 & 24 & 0 & 1 & -1 & 0 & -1 & 34 & 1 & 1 & -1 & -1 & -1 \\
5 & -1 & -1 & 1 & -1 & 1 & 15 & -1 & 1 & 0 & 0 & 0 & 25 & 1 & -1 & -1 & -1 & 1 & 35 & 1 & 1 & -1 & 1 & 1 \\
6 & -1 & -1 & 1 & 1 & -1 & 16 & -1 & 1 & 1 & -1 & -1 & 26 & 1 & -1 & -1 & -1 & 1 & 36 & 1 & 1 & -1 & 1 & 1 \\
7 & -1 & -1 & 1 & 1 & -1 & 17 & -1 & 1 & 1 & -1 & -1 & 27 & 1 & -1 & -1 & 1 & -1 & 37 & 1 & 1 & 1 & -1 & 1 \\
8 & -1 & 0 & -1 & 0 & 1 & 18 & -1 & 1 & 1 & 1 & 1 & 28 & 1 & -1 & -1 & 1 & -1 & 38 & 1 & 1 & 1 & -1 & 1 \\
9 & -1 & 0 & -1 & 0 & 1 & 19 & -1 & 1 & 1 & 1 & 1 & 29 & 1 & -1 & 1 & -1 & -1 & 39 & 1 & 1 & 1 & 1 & -1 \\
10 & -1 & 1 & -1 & -1 & 1 & 20 & 0 & -1 & -1 & 1 & 0 & 30 & 1 & -1 & 1 & -1 & -1 & 40 & 1 & 1 & 1 & 1 & -1
\end{tabular}
}}
\end{table}

%% file: appendix_trace_criteria.tex
\section{Trace-based Criteria}
\label{appendix_trace_criteria}

\textbf{Lack-of-fit criterion}

Together with the Lack-of-fit DP-criterion derived in Section \ref{section::LoF_criterion}, we formulate a criterion to estimate the lack-of-fit by minimizing the average squared lengths of posterior confidence intervals for linear functions of $\bm{\beta}_q$ defined by matrix $\bm{P}$. We define the ``Lack-of-fit LP-criterion'' as the mean of the squared lengths of the $100(1-\alpha_{LoF})\%$ posterior confidence intervals for these linear functions, i.e.\ we minimize
\begin{equation}
\label{eq::LoFLP_criterion}
\frac{1}{q}\mbox{trace}\left[\bm{PP}^T\left(\bm{L}+\frac{\bm{I}_{q}}{\tau^{2}}\right)^{-1}\right]F_{1,d;1-\alpha_{LoF}.}
\end{equation} 
This trace-based criterion is linked to the lack-of-fit part of Generalized $L$-optimality \citep{Goos2005model}, and the pure error estimation approach retains the corresponding upper point of the F-distribution. Henceforth we mainly consider the case when $\bm{PP}^T$ is diagonal, and the criterion above is reduced to weighted-$AP$-optimality. In other words, the ``Lack-of-fit  AP-criterion'' stands for minimization of the weighted average of the $q$-dimensional vector of the posterior confidence intervals' squared lengths for the potential parameters.

Maximizing the (weighted) trace of the dispersion matrix $\bm{L}$ translates into maximizing the (weighted) mean distance of potential terms from the linear subspace spanned by the primary terms. Aiming towards the primary and potential subspaces being as near to orthogonal to each other as possible also works towards maximizing the power of the lack-of-fit test. 

\textbf{MSE-based criterion}

To derive the trace-based form of the MSE criterion, we calculate the expectation of the trace function of the MSE matrix (\ref{eq::MSE}), under the prior for the potential terms $\bm{\beta}_q \sim \mathcal{N}(\bm{0},\tau^{2}\sigma^{2}\bm{I}_{q})$:
\begin{align*}
\mathtt{E}_{\bm{\beta}_q}\mbox{trace}[\mbox{MSE}(\bm{\hat{\beta}}_p|\bm{\beta}_q)]&=\mbox{trace}[\mathtt{E}_{\bm{\beta}_q}\mbox{MSE}(\bm{\hat{\beta}}_p|\bm{\beta}_q)]\\&=\mbox{trace}[\sigma^2(\bm{X}_p^{T}\bm{X}_p)^{-1} + \mathtt{E}_{\bm{\beta}_q}(\bm{A}\bm{\beta}_q\bm{\beta}_q^T\bm{A}^T)]\\&=\mbox{trace}[\sigma^2(\bm{X}_p^{T}\bm{X}_p)^{-1}+\sigma^2\tau^2\bm{A}\bm{A}^T]\\&=\sigma^2\mbox{trace}[(\bm{X}_p^{T}\bm{X}_p)^{-1}+\tau^2\bm{A}\bm{A}^T]\\&=\sigma^2[\mbox{trace}\{(\bm{X}_p^{T}\bm{X}_p)^{-1}\}+\tau^2\mbox{trace}\bm{A}\bm{A}^T].
\end{align*}

The operations of calculating trace and expectation are commutative, hence there is no necessity of any additional numerical evaluations, and in the case of the trace-based criterion using the point prior for $\bm{\beta}_q$ at $\bm{\beta}_q = \pm\sigma\tau\bm{1}_q$ would lead to the same resulting function. By minimizing the whole function above, we simultaneously minimize both the average variance of the primary terms and the expected squared norm of the bias vector in the direction of the potential terms, scaled by $\tau^2$ which regulates the magnitude of the potential terms relative to the error variance. We formally define the ``MSE(L)-criterion'' as being to minimize
\begin{equation}
\label{eq::MSE_L}
\frac{1}{p}\mbox{trace}\{(\bm{X}^T_{p}\bm{X}_{p})^{-1}+\tau^2\bm{A}\bm{A}^T\}.
\end{equation}

\textbf{Compound criterion}

Similarly, we obtain the trace-based ``compound MSE-LP$_S$-criterion'' by joining the LP$_S$ criterion with trace-based lack-of-fit (\ref{eq::LoFLP_criterion}) and MSE components to minimize
\begin{align}
\label{eq::MSE_LP}
&\left[\frac{1}{p-1}\mbox{trace}(\bm{WX}^T_{p-1}\bm{Q}_{0}\bm{X}_{p-1})^{-1}F_{1,d;1-\alpha_{LP}}\right]^{\kappa_{LP}}\times \notag\\& \left[\frac{1}{q}\mbox{trace}\left(\bm{L}+\frac{\bm{I}_{q}}{\tau^{2}}\right)^{-1}F_{1,d;1-\alpha_{LoF}}\right]^{\kappa_{LoF}}\times 
\\& \left[\frac{1}{p-1}\mbox{trace}[\bm{M}^{-1}+\tau^2\bm{A}\bm{A}^T]_{[p-1, p-1]}\right]_.^{\kappa_{MSE}} \notag
\end{align}
Here $[\bm{M}^{-1}+\tau^2\bm{A}\bm{A}^T]_{[p-1, p-1]}$ stands for the submatrix corresponding to the parameters of interest, that is with the first row and first column removed.
Confidence levels $\alpha_{LP}$ and $\alpha_{LoF}$ play similar roles here, although they do not have to be the same as in the determinant-based criterion. Moreover, it would be sensible to take into account the multiple testing corrections, as we are dealing with minimizing the lengths of multiple confidence intervals rather than with the volume of a single region. 

\subsection{Example}
\label{appendix_trace_example}

Table \ref{tab::MSE(L)_ex1} below provides a summary of MSE-LP$_S$-optimal designs (optimality criterion as in (\ref{eq::MSE_LP})) for the example considered in Section \ref{subsec::unblocked_example}. 

Each row corresponds to a design optimal according to the compound criterion with the combination of weights $\kappa_i$. The distribution of degrees of freedom between the pure error and lack-of-fit components in the designs and the optimal designs' efficiencies with respect to the individual criteria that are given in the columns. 

\begin{table}
    \caption{\label{tab::MSE(L)_ex1}Properties of $MSE-LP_{S}$-optimal designs}
    
\fbox{%
\resizebox{\textwidth}{!}{
\begin{tabular}{*{12}{r}}
& \multicolumn{3}{l}{\textbf{Criteria, $\bm{\tau^2=1}$}} & \multicolumn{2}{l}{\textbf{DoF}} & \multicolumn{6}{l}{\textbf{Efficiency,\%}}   \\
& \textbf{LP}       & \textbf{LoF(LP)}    & \textbf{MSE(L)}   & \textbf{PE}        & \textbf{LoF}        & \textbf{DP}   & \textbf{LoF(DP)}   & \textbf{MSE(D)}  &  \textbf{LP}       & \textbf{LoF(LP)}   & \textbf{MSE(L)}  \\ 
1 & 1    & 0    & 0    & \multicolumn{1}{|r}{16} & \multicolumn{1}{r|}{3} & 97.54 & 53.86 & 92.54 & \multicolumn{1}{|r}{100.00} & 96.87  & 12.08  \\
2 & 0    & 1    & 0    & \multicolumn{1}{|r}{13} & \multicolumn{1}{r|}{6} & 35.43 & 81.99 & 36.72 & \multicolumn{1}{|r}{0.00}   & 100.00 & 0.00   \\
3 & 0    & 0    & 1    & \multicolumn{1}{|r}{4} & \multicolumn{1}{r|}{15} & 18.67 & 38.43 & 51.84 & \multicolumn{1}{|r}{11.73}  & 34.79  & 100.00 \\
4 & 0.5  & 0.5  & 0    & \multicolumn{1}{|r}{15} & \multicolumn{1}{r|}{4} & 95.14 & 60.37 & 92.78 & \multicolumn{1}{|r}{99.80}  & 98.12  & 13.99  \\
5 & 0.5  & 0    & 0.5  & \multicolumn{1}{|r}{12} & \multicolumn{1}{r|}{7} & 77.77 & 72.05 & 84.91 & \multicolumn{1}{|r}{81.10}  & 98.72  & 25.19  \\
6 & 0    & 0.5  & 0.5  & \multicolumn{1}{|r}{9} & \multicolumn{1}{r|}{10} & 36.80 & 70.91 & 51.12 & \multicolumn{1}{|r}{28.13}  & 91.60  & 83.52  \\
7 & 1/3  & 1/3  & 1/3  & \multicolumn{1}{|r}{11} & \multicolumn{1}{r|}{8} & 69.53 & 73.16 & 79.71 & \multicolumn{1}{|r}{70.59}  & 97.47  & 27.98  \\
8 & 0.5  & 0.25 & 0.25 & \multicolumn{1}{|r}{12} & \multicolumn{1}{r|}{7} & 77.20 & 72.83 & 84.44 & \multicolumn{1}{|r}{81.47}  & 98.80  & 23.88  \\
9 & 0.25 & 0.5  & 0.25 & \multicolumn{1}{|r}{12} & \multicolumn{1}{r|}{7} & 70.90 & 69.80 & 78.15 & \multicolumn{1}{|r}{72.16}  & 98.49  & 26.19  \\
10 & 0.25 & 0.25 & 0.5 & \multicolumn{1}{|r}{9} & \multicolumn{1}{r|}{10} &  41.12 & 75.50  & 56.37  & \multicolumn{1}{|r}{34.49} & 92.47 & 77.27\\ 
 \hline
& & & & & & & & & & & \\
 & \multicolumn{3}{l}{\textbf{Criteria, $\bm{\tau^2=1/q}$}} & \multicolumn{2}{l}{\textbf{DoF}} & \multicolumn{6}{l}{\textbf{Efficiency,\%}} \\
  & \textbf{LP}       & \textbf{LoF(LP)}    & \textbf{MSE(L)}   & \textbf{PE}        & \textbf{LoF}        & \textbf{DP}   & \textbf{LoF(DP)}   & \textbf{MSE(D)}  & \textbf{LP}       & \textbf{LoF(LP)}   & \textbf{MSE(L)}  \\
1 & 1    & 0    & 0    & \multicolumn{1}{|r}{16} & \multicolumn{1}{r|}{3} & 97.54 & 92.13 & 92.18 & \multicolumn{1}{|r}{100.00} & 95.61  & 52.55  \\
2 & 0    & 1    & 0    & \multicolumn{1}{|r}{16} & \multicolumn{1}{r|}{3} & 39.66 & 100.00 & 37.95 & \multicolumn{1}{|r}{0.13}   & 100.00 & 0.12   \\
3 & 0    & 0    & 1    & \multicolumn{1}{|r}{3} & \multicolumn{1}{r|}{16} & 0.77  & 0.93  & 83.48 & \multicolumn{1}{|r}{0.02}   & 0.01   & 100.00 \\
4 & 0.5  & 0.5  & 0    & \multicolumn{1}{|r}{17} & \multicolumn{1}{r|}{2} & 96.87 & 94.29 & 89.84 & \multicolumn{1}{|r}{97.97}  & 97.80  & 51.17  \\
5 & 0.5  & 0    & 0.5  & \multicolumn{1}{|r}{12} & \multicolumn{1}{r|}{7} & 79.66 & 87.18 & 86.32 & \multicolumn{1}{|r}{84.81}  & 88.22  & 79.60  \\
6 & 0    & 0.5  & 0.5  & \multicolumn{1}{|r}{13} & \multicolumn{1}{r|}{6} & 76.46 & 89.87 & 80.78 & \multicolumn{1}{|r}{79.23}  & 91.48  & 79.11  \\
7 & 1/3  & 1/3  & 1/3  & \multicolumn{1}{|r}{13} & \multicolumn{1}{r|}{6} & 81.53 & 90.09 & 85.52 & \multicolumn{1}{|r}{85.96}  & 91.56  & 76.65  \\
8 & 0.5  & 0.25 & 0.25 & \multicolumn{1}{|r}{15} & \multicolumn{1}{r|}{4} & 90.60 & 91.94 & 88.43 & \multicolumn{1}{|r}{95.09}  & 94.75  & 63.82  \\
9 & 0.25 & 0.5  & 0.25 & \multicolumn{1}{|r}{15} & \multicolumn{1}{r|}{4} & 84.12 & 92.88 & 83.39 & \multicolumn{1}{|r}{87.86}  & 95.51  & 72.66  \\
10 & 0.25 & 0.25 & 0.5 & \multicolumn{1}{|r}{13} & \multicolumn{1}{r|}{6} & 77.72 & 89.15 & 82.70 & \multicolumn{1}{|r}{82.34} & 90.96 & 81.76\\
\\
\end{tabular}
}}
\end{table}

In general, the designs tend to be quite $DP$- and $LP$-efficient. $DP$-efficient designs (Table \ref{tab::MSE(D)_ex1}) are not bad in terms of $LP$-efficiency and vice versa, but the same cannot be observed for the lack-of-fit components and seems not to be true at all for the $MSE$ components, especially, for the $MSE(L)$-optimal design when $\tau^2=1$.

$MSE-LP_{S}$-optimal designs tend to have larger $LP$- and $MSE(L)$-efficiencies in the case of smaller $\tau^2$, which makes sense -- smaller potential contamination leads to a more easily achievable compromise between the contradicting components of the criteria (the same is observed for the trace-based efficiencies of the $MSE-DP_{S}$-optimal designs). It is also notable that the $LoF(LP)$-optimal design is also $LoF(DP)$-optimal for $\tau^2=1/q$.

The $MSE(L)$-component seems to be much more sensitive to the weight allocations than the $MSE(D)$ component: in the case of $\tau^2=1$ reasonable efficiencies are achieved only when most of the weight is on the `potential terms' criterion components, i.e.~designs \#$3$, \#$6$ and \#$10$. 

\subsection{Blocked Experiments}

Following the derivations for determinant-based criteria in Section \ref{sec::mse_blocked}, we take the
expectation of the trace of (\ref{eq::mesb_submatrix}) to obtain the trace-based MSE-criterion:
\begin{align}
\label{eq::MSE_B_tr}
\mathtt{E}_{\beta_q}\mbox{trace}[\mbox{MSE}(\bm{\hat{\tilde{\beta}}}_p|\bm{\tilde{\beta}})_{pp}]&= \mbox{trace}[\mathtt{E}_{\beta_q}\mbox{MSE}(\bm{\hat{\tilde{\beta}}}_p|\bm{\tilde{\beta}})_{pp}]\notag \\&=\mbox{trace}[\sigma^2\bm{\tilde{M}}^{-1}_{pp}+\mathtt{E}_{\beta_q}(\bm{\tilde{A}}\bm{\beta}_q\bm{\beta}_q^T\bm{\tilde{A}})_{pp}]\notag\\&=  \sigma^2\mbox{trace}[\bm{\tilde{M}}^{-1}_{pp}+\tau^2\{\bm{\tilde{A}}\bm{\tilde{A}}^T\}_{pp}]\notag \\&=\sigma^2[\mbox{trace}(\bm{X}^T_{p}\bm{QX}_{p})^{-1}+\tau^2\mbox{trace}\{\bm{\tilde{A}}\bm{\tilde{A}}^T\}_{pp}].
\end{align}

The MSE-LP compound criterion for a blocked experiment is then to minimize
\begin{align}
\label{eq::MSE_L_B}
&\left[\frac{1}{p}\mbox{trace}(\bm{WX}^T_{p}\bm{Q}\bm{X}_{p})^{-1}F_{1,d_B;1-\alpha_{LP}}\right]^{\kappa_{LP}}\times \notag \\&\left[\frac{1}{q}\mbox{trace}\left(\bm{\tilde{L}}+\bm{I}_{q}/\tau^{2}\right)^{-1}F_{1,d_B;1-\alpha_{LoF}}\right]^{\kappa_{LoF}}\times \notag \\& \left[\frac{1}{p}\mbox{trace}\{(\bm{X}^T_{p}\bm{QX}_{p})^{-1}+\tau^2[\bm{\tilde{A}}\bm{\tilde{A}}^T]_{pp}\}\right]^{\kappa_{MSE}}.
\end{align}

%% file: appendix_case_study.tex
\section{Case-study, blocked experiment}
\label{appendix_case_study}
MSE-DP-optimal blocked design for the case study presented in Section \ref{subsec::case_study}.

\begin{table}
    \caption{\label{tab::CS_Design}Case-study: $MSE-DP$-optimal design \#$1$ with two center points, $\tau^2=1$}
\centering    
\scalebox{0.75}{
\fbox{%
\begin{tabular}{rrrrrrrr|r|rrrrrrrr}
\multicolumn{1}{l}{} & \multicolumn{7}{l|}{Block I} & \multicolumn{1}{l|}{} & \multicolumn{1}{l}{} & \multicolumn{7}{l}{Block II}\\ \hline 
\multicolumn{1}{l}{} & \multicolumn{1}{l}{X1} & \multicolumn{1}{l}{X2} & \multicolumn{1}{l}{X3} & \multicolumn{1}{l}{} & \multicolumn{1}{l}{X1} & \multicolumn{1}{l}{X2} & \multicolumn{1}{l|}{X3} & \multicolumn{1}{l|}{} & \multicolumn{1}{l}{} & \multicolumn{1}{l}{X1} & \multicolumn{1}{l}{X2} & \multicolumn{1}{l}{X3} & \multicolumn{1}{l}{} & \multicolumn{1}{l}{X1} & \multicolumn{1}{l}{X2} & \multicolumn{1}{l}{X3} \\ \hline
1 & -1 & -1 & -1 & 10 & 0 & 0 & 0 &  & 1 & -1 & -1 & -1 & 10 & 0 & 0 & 0 \\
2 & -1 & -1 & 0 & 11 & 0 & 1 & -1 &  & 2 & -1 & -1 & 0 & 11 & 0 & 1 & 1 \\
3 & -1 & -1 & 1 & 12 & 1 & -1 & -1 &  & 3 & -1 & -1 & 1 & 12 & 1 & -1 & -1 \\
4 & -1 & 0 & -1 & 13 & 1 & -1 & 0 &  & 4 & -1 & 0 & 1 & 13 & 1 & -1 & 0 \\
5 & -1 & 1 & -1 & 14 & 1 & -1 & 1 &  & 5 & -1 & 1 & -1 & 14 & 1 & -1 & 1 \\
6 & -1 & 1 & 1 & 15 & 1 & 0 & 1 &  & 6 & -1 & 1 & 0 & 15 & 1 & 0 & -1 \\
7 & -1 & 1 & 1 & 16 & 1 & 1 & -1 &  & 7 & 0 & -1 & -1 & 16 & 1 & 0 & 1 \\
8 & 0 & -1 & 1 & 17 & 1 & 1 & 0 &  & 8 & 0 & -1 & 1 & 17 & 1 & 1 & -1 \\
9 & 0 & 0 & 0 & 18 & 1 & 1 & 1 &  & 9 & 0 & 0 & 0 & 18 & 1 & 1 & 1
\end{tabular}
}}
\end{table}

\begin{figure}[!h]
\centering
\makebox{\includegraphics[scale=0.65]{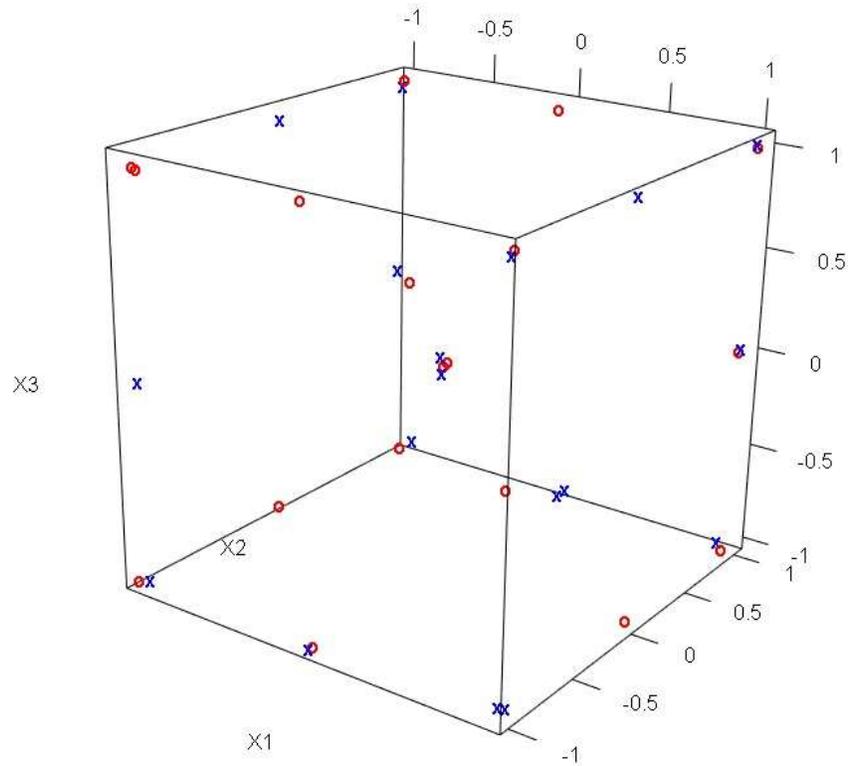}}    
\caption{\label{fig::CS_design}$MSE-DP$-optimal design \#$1$: colours (blue and red) and symbols (`x' and `o') serve as block indicators.}
\end{figure} 